  \def\real{{\mathchoice
    {\hbox{$\displaystyle\kern-.2mm 1\kern-1mm \mbox{\rm R}\kern-.2mm$}}
    {\hbox{$\textstyle\kern-.2mm 1\kern-1mm \mbox{\rm R}\kern-.2mm$}}
    {\hbox{$\scriptstyle\kern-.2mm 1\kern-1mm \mbox{\rm R}\kern-.2mm$}}
    {\hbox{$\scriptscriptstyle \kern-.2mm 1\kern-1mm \mbox{\rm R}\kern-.2mm$}}}}

\documentstyle[12pt,twoside]{article}

\begin{document}
\setcounter{page}{1}
\title{The Determination of the Orbit Spaces of Compact Coregular Linear Groups}

\author{V. Talamini}
\date{INFN, Sezione di Padova\\I-35131 Padova, Italy\\
 (e-mail: talamini@padova.infn.it)}
\markboth{V. Talamini}{The Determination of the Orbit Spaces of
 Compact Coregular Linear Groups}
\maketitle

\begin{abstract}
Some aspects of phase transitions can be more conveniently studied in the
orbit space of the action of the symmetry group. After a brief review of
the fundamental ideas of this approach, I shall concentrate on the
mathematical aspect and more exactly on the determination of the equations
defining the orbit space and its strata. I shall deal only with compact
coregular linear groups.
The method exposed has been worked out together with prof. G. Sartori and
it is based on the solution of a matrix differential equation.
Such equation is easily solved if an integrity basis of the
group is known. If the integrity basis is unknown one may determine
 anyway for which degrees of the basic invariants there are solutions to the
equation, and in all these cases also find out the explicit form of the
 solutions.
 The solutions determine completely the stratification of the orbit spaces.
Such calculations have been carried out for 2, 3 and 4-dimensonal orbit
spaces.
The method is of general validity but the complexity of the calculations
rises tremendously with the dimension $q$ of the orbit space.
Some induction rules have been found as well. They allow to determine
easily most of the solutions for the $(q+1)$-dimensional case once
the solutions for the $q$-dimensional case are known.
The method exposed is interesting because it allows to determine the orbit
 spaces without using any specific knowledge of group structure and integrity
 basis and evidences a certain hidden and yet unknown link
 with group theory and invariant theory.
\end{abstract}

\section{Introduction}
The symmetry of a given physical system is represented mathematically by a
transformation group $G$ that acts on a representation
 space $X$. One has also to deal with functions defined on $X$ that remain
 invariant with respect to $G$-transformations. In all such systems
 it is clear that in all points of $X$ connected by $G$-transformations,
 the values assumed by the invariant functions do not change, that is the
 invariant functions are constant on the orbits of $X$. When one wants to
 study the variations of the values $f(x)$ taken by a $G$-invariant function
 $f$ (for example in the study of the extrema of $f$),
 one might limit to consider variations of $x$ from one orbit to another.
This fact leads to prefer the description of the physical system in the orbit
 space of the symmetry group.
 In the orbit space, in fact, a whole orbit of $X$ is shrinked
 to a single point and all invariant functions can be considered
 as functions defined on the orbit space.\\
Another important advantage introduced by the orbit space
 description of the physical system concerns phase transitions.
The points $x \in X$ are transformed by the action of the group $G$ to all
 the points of the orbit through $x$, however,
 the isotropy subgroup $G_x$ of $x$ leaves $x$ fixed.
If $x$, minimum point of a $G$-invariant potential, represents
 a stable configuration for the physical system, then
 the true physical symmetry is determined by the isotropy subgroup $G_x$ of $x$.
A stratum contains all the points of $X$ whose isotropy subgroups are
 conjugated in $G$. When a continuous phase transition takes place the
 physical symmetry $G_x$ changes to a subgroup or a supergroup of $G_x$.
Consequently, the minimum point of the potential changes its position with
 continuity from one stratum to another one.
In the orbit space the strata are reduced to their essential form, having
 shrinked orbits to single points, but all the physically relevant properties
 are inherited because the stratification of the orbit space represents
 all the phases allowed for the physical system in an essential way.\\
To carry out all these ideas, one has to give a precise
 mathematical description of the orbit spaces.
Following Gufan \cite{1guf} one can employ the integrity basis to parametrize
 the orbits in the orbit space.
If the symmetry group is a compact group, the orbit space can be
 identified with a closed connected semialgebraic subset $S$ of $\real^q$,
 with $q$ the finite number of the basic homogeneous polynomials of the
 integrity basis. The dimension of $S$ is $q$ if and only if the
 group $G$ is coregular, that is if and only if the basic
 invariants are algebraic independent
 \cite{2as,3ps}. This will be the only case discussed here (however, some
 extensions to non coregular compact groups have been obtained too \cite{4sv}).
All the boundary of $S$ is contained in the set of the zeros of a
 homogeneous polynomial $A$ which is a factor of the determinant of a
 $q \times q$ matrix $P$ constructed out of the gradients of the basic
invariants.
The matrix $P$ and the polynomial $A$
 satisfy a differential matrix equation (here called the boundary equation)
 and some very general initial conditions \cite{5s,6st}. These initial
 conditions are necessarily satisfied if $P$ is determined by an integrity
 basis of a compact coregular group.
If the integrity basis is known one can write down easily $P$ and $A$ and
 determinine the set $S$ completely. However, the integrity basis of a
general compact coregular linear group is not known.
We tried to overcome this difficulty reversing the problem and looking for all
 solutions $P$ of the boundary equation satisfying the initial conditions. In
 doing this we only have to fix the dimension $q$ of $P$.
 The initial conditions imposed are so strong that the boundary
 equation can be solved analytically and it turns out that for likely any
 value of $q$ it has only a finite number of different solutions.
 \cite{6st,7st}.\\
The method employed to determine and classify the $P$-matrices is of general
 validity and can be extended to higher and higher dimensions but the
 complexity of the calculations rises tremendously.
Some inductive rules have been found as well. These allow to determine
 easily many of the solutions for the $(q+1)$-dimensional case once the
 solutions for the $q$-dimensional case are known \cite{10st}.\\
It is not clear if all solutions $P$ of the boundary equation are also
 determined from an integrity basis of an existing group. However, the
 converse is true, that is all compact coregular
 groups must determine $P$-matrices that satisfy the boundary equation.
One then may classify all possible orbit spaces.\\
In what follows I shall present briefly, but in more detail, the ideas just
 sketched but I shall skip all the proofs.
 The interested reader is referred to
 \cite{8b,9s,mic} and references therein for general properties of compact
 group actions and to \cite{6st,10st} for technical details about our
 statements.

\section{Group Actions}

Let $G$ be any compact group of transformations of a finite dimensional
 space. In all generality, we can suppose $G$ as a group of $n \times n$
 orthogonal real matrices acting in the Euclidean space $\real^n$.\\
The {\it orbit} through a point $x$ is the set of points obtained
 from $x$ by means of $G$-transformations:
$$\Omega(x) = \{g \cdot x,\quad \forall g \in G\}, \qquad  x\in\real^n$$
The {\it isotropy subgroup} $G_x$ of the point $x$ is the subgroup of $G$ that
 leaves $x$ fixed:
$$G_x = \{g \in G \mid g \cdot x = x\}, \qquad  x\in\real^n$$
Points lying in a same orbit have conjugated isotropy subgroups, in fact
 the following relation holds true:
$$G_{g \cdot x} = g \cdot G_x \cdot g^{-1},  \qquad g\in G,\quad x\in\real^n$$
So, to each orbit $\Omega(x)$ of $\real^n$ one can associate the conjugacy
 class of its isotropy subgroups $[G_x]$, called the {\it orbit type} of
 $\Omega(x)$.
 I will sometimes use the term {\it symmetry} in place of orbit type to
 remind its physical meaning.
 Orbits can then be classified in terms of their symmetry.\\
In general there are many orbits of the same symmetry.\\
 A {\it stratum} is the subset of $\real^n$
 consisting of all the orbits of the same symmetry,
 i. e. of all the points whose isotropy subgroups are conjugated in $G$:
$$\Sigma_{[H]} = \{ x\in\real^n \mid G_x \in [H] \}$$
The stratum $\Sigma_{[H]}$, containing orbits of symmetry $[H]$, is said of
 {\it symmetry} $[H]$.
 Clearly there is a one to one correspondence between orbit types and strata.
Each point $x$ of $\real^n$ belongs to only one stratum, that one of symmetry
 $[G_x]$, so $\real^n$ is the disjoint union of all its strata and one says that
 $\real^n$ is {\it stratified} by the group action.\\
One can introduce a partial ordering relation in the set of orbit types.
The orbit type $[K]$ being {\it greater}
 of the orbit type $[H]$ if the groups in $[H]$ are subgroups of
 the groups in $[K]$. This relation is only partial because,
 given two arbitrary orbit types, it is not always true that the groups in one class
 are subgroups of the groups in the other class.
The greatest orbit type is always $[G]$ and its stratum consists of all fixed
 points of the $G$-action. It contains at least the origin of $\real^n$.\\
Two important facts are true for compact groups:
\begin{enumerate}
  \item The number of orbit types is finite;
  \item There is a unique minimal orbit type, the {\it principal type}.
\end{enumerate}
One then has a finite stratification of $\real^n$ with a unique
{\it principal stratum} of minimal symmetry and several {\it singular strata}
of greater symmetry.\\
The {\it orbit space (OS)} of the action of $G$ in $\real^n$ is the quotient
 space $\real^n/G$.\\
The projection:
$$\pi : \real^n \to \real^n/G$$
maps the orbits of $\real^n$ into single points of the OS.\\
Projections of strata of $\real^n$ form {\it strata of the OS}.\\
The projection of the principal stratum of $\real^n$ forms the {\it principal
 stratum of the OS}. It is always an open, connected, dense subset of the
 OS (even if the group is not connected).\\
If $\Sigma_{[K]}$ is a stratum of $\real^n$ of greater symmetry
than $\Sigma_{[H]}$,
 then $\pi(\Sigma_{[K]})$ lies in the boundary of $\pi(\Sigma_{[H]})$.
 As a consequence the boundary
 of the principal stratum contains all singular strata.\\
Clearly there is a one to one correspondence between the strata of $\real^n$
 and the strata of the OS. So the OS is stratified in exactly the same
 manner as $\real^n$ is.

\section{Orbit Spaces in $\real^q$}

An explicit mathematical description of the OS is obtained by means of
 the {\it integrity basis (IB)} of the group.\\
For compact groups the ring of all real $G$-invariant
 polynomial functions admits a finite integrity basis \cite{11h,12n}:
$$p_1(x), p_2(x), \ldots, p_q(x), \qquad  p_i(x)\in\real^n[x]^G$$
All real $G$-invariant polynomial functions
$f(x),\ x\in\real^n$, can be expressed as real polynomials $\hat f(p)
,\ p\in\real^q$, of the $q$ basic invariants $p_i(x)$:
$$ \forall f\in\real[x]^G,\ \exists \hat f\in\real[p] \
\mid\ f(x) = \hat f(p_1(x),\ldots,p_q(x)) $$
A similar theorem also holds for real $G$-invariant $C^\infty$-functions
 \cite{13s}.\\
All real $G$-invariant $C^\infty$-functions
$f(x),\ x\in\real^n$, can be expressed as real $C^\infty$-functions $\hat f(p)
,\ p\in\real^q$, of the $q$ basic invariants $p_i(x)$.\\
The IB is supposed minimal, i. e. no element of the IB can be expressed as a
polynomial in the other ones. \\
The basic invariant polynomials $p_1(x),\ldots,p_q(x)$ can always be chosen
 homogeneous.\\
 If there are no fixed points except the origin, as it is here supposed, the
 lowest degree of the $p_i(x)$ is two. (If there are some non zero fixed
 points the representation is reducible and one may consider the restriction
 of the $G$-action to the invariant subspace in which $G$ acts effectively).\\
One may order the polynomials of the IB according to their
 degrees:
$$  d_1 \geq d_2 \geq \cdots \geq d_q = 2 $$
For the orthogonality of $G$, the form of the last invariant $p_q(x)$ is
 always the following:
$$ p_q(x) = (x,x) = \sum_{i=1}^n x_i^2  $$
The group $G$ is said {\it coregular}
if the basic polynomials are algebraically independent, that is if no
polynomial $f$ such that:
$$ f(p_1(x),\ldots,p_q(x)) = 0 \qquad   \forall x \in \real^n$$
exists. Otherwise $G$ is said {\it non coregular}.\\
The choice of the IB is not unique, but the group fixes the number
 $q$ of its elements and their degrees $d_1,\ldots,d_q$.\\
All {\it IB transformations (IBTs)}:
$$p_i'(x) =  p_i'(p_1(x),\ldots,p_q(x))$$
must have Jacobian matrix $J_{ij}(x)=\partial p_i'(x)/\partial p_j(x)$
which is an upper triangular (or block triangular) matrix with
elements that are $G$-invariant homogeneous polynomials with degrees:
$$\deg(J_{ij}(x)) = d_i - d_j $$
Then, $J_{ij}(x)$ is identically zero if $di - dj$ is not a linear
 combination of the degrees $d_1,\ldots,d_q$ of the basic invariants.
The particular form of $J(x)$ implies that its diagonal blocks and its
determinant are non vanishing constants.\\
The invariant polynomials separate the orbits:
$$\forall\ \Omega_1 \neq \Omega_2,\ \exists\ f\in\real[x]^G \mid
      f(x_1) \neq f(x_2),\ \forall x_1\in\Omega_1,\ x_2\in\Omega_2$$
Because $f(x)$ can be expressed as $\hat f(p_1(x),\ldots,p_q(x))$ the IB
 separates the orbits:
$$\forall\ \Omega_1 \neq \Omega_2,\ \exists\ p_i\in\mbox{IB} \mid
      p_i(x_1) \neq p_i(x_2),\ \forall x_1\in\Omega_1,\ x_2\in\Omega_2$$
Then the IB can be used to represent orbits of $\real^n$ as points
 of $\real^q$, in fact
$\forall x \in \Omega,\mbox{the vector } (p_1(x), p_2(x),\ldots, p_q(x))\ $
is constant. The $q$ numbers $p_i=p_i(x),\ x \in \Omega$, determine
 the point  $p=(p_1,p_2,\ldots,p_q)\in\real^q$,
which can be considered the image in $\real^q$ of the orbit
 $\Omega\subset\real^n$.
No other orbit of $\real^n$ is represented in $\real^q$ by the same point.\\
The map:
$$p:\real^n \to \real^q:x \to (p_1(x), p_2(x),\ldots, p_q(x))$$
is called the {\it orbit map}.\\
The orbit map induces a one to one correspondence between $\real^n/G$ and
 the subset
$$S=p(\real^n) \subset\real^q,$$
So, $S$ can be identified with the OS of the $G$-action. \\
$S$ is a closed connected subset of $\real^q$ and it has dimension $q$ if
 $G$ is coregular. If $G$ is non coregular, $S$ is contained in the surface
 defined by the algebraic relations between the polynomials of the IB.\\
The strata of $S$ are the images of the strata of $\real^n$ through the orbit
map. The principal stratum lies in the interior of $S$ and its boundary
 contains all singular strata.\\
The origin $O$ of $\real^n$ always has for image the origin $O$ of $\real^q$,
 because of the homogeneity of the IB.\\
$S$ is strictly contained in $\real^q$.
As a trivial example one may consider the rotation group of the plane, where
the only invariant is $p_1(x,y)=x^2+y^2$
and the OS is $S=\{p_1 \in \real \mid p_1\geq 0\}$.
So, one must determine explicitly the
 equations and inequalities defining $S$ as a subset of $\real^q$\\
The points $x$ and $cx$, $x \in \real^n$, $c\in\real$, $c\ne 0$,
 always belong to
 the same stratum of $\real^n$ because the linearity of the $G$-action
 implies $G_x=G_{cx}$.
All strata but the origin have then an intersection with the hypersphere
 of radius 1, defined by the equation $(x,x)=1$.
All strata of $S$ but the origin have then an intersection with the
 hyperplane $\Pi=\{p\in\real^q \mid p_q=1\}$.
The intersection $S\cap\Pi$ gives a $(q-1)$-dimensional compact connected
 image of $S$ in $\Pi$ containing all strata except the origin.
Varying $p_q$ this section varies its size and becomes smaller and smaller
 approaching the origin of $\real^q$ but it mantains its geometrical shape.
 So these sections are suitable to give a clearer image of the orbit space.\\
The $G$-invariant functions are constant on the orbits:
$$ f(g \cdot x) = f(x)  \qquad \forall g\in G,\ x\in\real^n$$
Therefore they can be thought as functions defined in the OS.
 In so doing one eliminates the
 degeneracy of all points in a same orbit in which the values assumed by
 all $G$-invariant functions are constant.\\
 As $G$-invariant functions can be expressed as functions of the basic
 invariants, and as $(p_1(x),\ldots,p_q(x))$ does not
 change its value along the orbits of $\real^n$, all $G$-invariant functions
 can be considered as functions of the variables $p_1,\ldots,p_q$ of $\real^q$:
$$f(x) = \hat f(p_1(x),\ldots,p_q(x))  \to \hat f(p_1,\ldots,p_q)$$
$\hat f(p)$ may be defined also in points $p\notin S$ but only the
restriction $\hat f(p)\mid_{p\in S}$ has the same range as $f(x),\
 x \in \real^n$.
The values assumed by $G$-invariant functions $f(x),\ x\in\real^n$ (in
 particular maxima or minima) are the values assumed by the
 corresponding functions $\hat f(p),\ p\in S$.\\
 From now on I shall drop the hat on the functions defined in $\real^q$.\\
A polynomial $f(p)$ is said {\it $w$-homogeneous} of {\it weight}
$d$ if $f(p(x))$ is homogeneous with degree $d$.
Each coordinate $p_1,\ldots,p_q$ of
$\real^q$ has then a weight $d_1,\ldots,d_q$.\\
The IBTs:
$$p_i'(x) = p_i'(p_1(x),\ldots,p_q(x)) $$
can be viewed as coordinate transformations of $\real^q$:
$$p_i' = p_i'(p_1,\ldots,p_q) $$
The Jacobian matrix $J_{ij}(p)=\partial p_i'(p)/\partial p_j$ is upper
triangular (or block triangular) as $J(x)$ is, with the
 elements $J_{ij}(p)$ $w$-homogeneous polynomials of weight $d_i-d_j$ and
 identically zero if $d_i-d_j$ is not a linear combination of the $d_i$'s.\\
The only coordinate transformations of $\real^q$ of our interest are those
 corresponding to IBTs. We call them {\it IBTs (of} $\real^q${\it)}.\\
The IBTs change the form of $S$ but not its topological structure
 and stratification because these depend on the group action and are
 independent from the basis.\\
In a point $x\in\real^n$, the number of linear independent gradients of the
 basic invariants depends on the symmetry of the stratum in which $x$ lies.
 One can use this fact to determine the boundary of $S$.
To this end it is convenient to construct the $q \times q$ grammian
matrix $P(x)$ with elements $P_{ab}(x)$ defined as follows:
$$   P_{ab}(x) = (\nabla p_a(x),\nabla p_b(x))$$
The matrix $P(x)$ has the following important properties:
\begin{enumerate}
\item $P(x)$ is a real, symmetric and positive semidefinite $q \times q$
matrix;
\item the matrix elements $P_{ab}(x)$ are $G$-invariant homogeneous
 polynomial functions of degree $d_a+d_b-2$ and the last row and column
 of $P(x)$ have the form:
$$ P_{aq}(x) = 2 d_a p_a(x)$$
\item $\mbox{rank}(P(x))$ is constant along the strata of $\real^n$;
\item $P(x)$ transforms as a contravariant tensor under IBTs:
$$     P_{ab}'(x) = J_{ai}(x) J_{bj}(x) P_{ij}(x) $$
\end{enumerate}
The $G$-invariance stressed at item 2. is due to the covariance of
 the gradients of the $G$-invariant functions
 ($\nabla f(g \cdot x)=g \cdot \nabla f(x)$)
 and to the orthogonality of $G$ which implies the invariance of
 the scalar products.\\
A very important consequence of item 2. is that all matrix elements of
 $P(x)$ can be expressed in terms of the basic invariants. One
 can then define a matrix $P(p)$ in $\real^q$ such that:
$$    P_{ab}(p(x)) = P_{ab}(x) \quad    \forall x \in \real^n$$
At the point $p=p(x)$, image in $\real^q$ of the point $x\in\real^n$
through the orbit map, the matrix $P(p)$ is the same as the matrix $P(x)$.
The matrix $P(p)$ is also defined outside $S$ but only in $S$ it has the
 same range as $P(x)$ in $\real^n$.\\
The properties of the matrix $P(p)$ depend on those of $P(x)$ and are the
 following:
\begin{enumerate}
\item $P(p)$ is a real, symmetric $q \times q$ matrix, which is positive
 semidefinite {\it only} in $S$;
\item the matrix elements $P_{ab}(p)$ are $w$-homogeneous polynomial
 functions of weight $d_a+d_b-2$ and the last row and column of $P(p)$
 have the form:
$$  P_{aq}(p) = 2 d_a p_a$$
\item $\mbox{rank}(P(p))$ is constant along the strata of $S$ and equals
the dimension of the stratum containing $p$;
\item $P(p)$ transforms as a contravariant tensor under IBTs:
$$  P_{ab}'(p) = J_{ai}(p) J_{bj}(p) P_{ij}(p)   $$
\end{enumerate}
Items 1. and 3. suggest how to find out the equations and inequalities
 defining $S$ and its strata:
$$S=\{p\in\real^q\mid P(p)\geq 0\}$$
$$S_k=\{p\in\real^q\mid P(p)\geq 0,\ \mbox{rank}(P(p))=k\}$$
where $S_k$ is the union of all $k$-dimensional strata.\\
If $G$ is coregular there is an open region
$S_q=\{p\in\real^q\mid P(p)>0,\ \mbox{rank}(P(p))=q\}$
 corresponding to the $q$-dimensional principal stratum.\\
 If $G$ is non-coregular the gradients of the IB are linearly dependent and
 $S$ is of lower dimension. In this case the whole of $S$ is contained in the
 surface determined by $\det(P(p))=0$.\\
The set $S$ is {\it semialgebraic} because it is defined by polynomial
equations and inequalities.\\
It is clear that the matrix $P(p)$ contains all information necessary to
 determine the orbit space $S$. So, in order to classify orbit spaces,
 it is sufficient to classify the corresponding matrices $P(p)$.

\section{The Boundary Equation}

To find out the equation of the boundary of $S$ in the case of coregular
 groups it is very useful to employ a matrix differential equation relating the
 matrix $P(p)$ and a polynomial $A(p)$ vanishing in all the boundary of $S$.\\
In some IB this {\it boundary equation}
has the following form:
$$P_{ab}(p) \partial_b A(p) =
  \sum_{b=1}^q P_{ab}(p) \partial/{\partial p_b} A(p) = 0 \qquad
                             a=1,\ldots,q-1$$
The polynomial $A(p)$ is a $w$-homogeneous factor of $\det(P(p))$,
called the {\it complete factor}.
The boundary of $S$ is exactly the region where $A(p)=0$ and $P(p)$ is
 positive semidefinite.\\
All irreducible factors of $\det(P(p))$ and all their products, satisfy
 the boundary equation too, but in different bases. The bases in which the
 $w$-homogeneous polynomial $a(p)$ satisfies the boundary equation are called
 {\it $a$-bases}. A polynomial that in some basis satisfies the boundary
 equation is called {\it active}.
 All irreducible factors of active polynomials are factors of $\det(P(p))$
 and the complete factor $A(p)$
 is the product of all irreducible active factors of $\det(P(p))$.
 The other factor $B(p)$ of $\det(P(p))$ is
 prime with respect to $A(p)$, it does not satisfy the boundary equation
 and it is called {\it passive}.
 $A(p)$ and $B(p)$ are uniquely defined except by non-zero
 constant factors.\\
We studied the properties of the boundary equation carefully, especially in
 connection with IBTs, and our main results are the following:
\begin{enumerate}

\item Given any active polynomial $a(p)$, all $a$-bases are
connected by IBTs not depending on $p_q$.
\item In all $A$-bases, the section $S \cap \Pi$ of $S$ with the hyperplane
 $\Pi=\{p \in \real^q \mid p_q=1\}$ contains the point
 $p_0=(0,\ldots,0,1)$ (the origin of $\Pi$) in its interior, and in $p_0$
 the restriction
 $A(p)\mid_{p\in\Pi}$ has its unique local non-degenerate extremum (positive
 by convention).
\item In all $A$-bases, $P(p_0)$ and the hessian $H(p_0)$
 ($H_{ab}(p) = \partial_a \partial_b A(p)$) of $A(p)$, evaluated at $p_0$, are
 diagonal (and closely related).
\end{enumerate}
These last facts imply that the weight $w(A)$ of $A(p)$ is bounded:
  $2 d_1 \leq w(A)\leq w(\det(P)) = 2 \sum_{i=1}^q (d_i-1)$, that
 in $A(p)$ there are no linear terms in
 $p_i,\ \forall i\leq q$, and that in $A(p)\mid_{p\in\Pi}$
the only quadratic terms are those in $\ p_i^2,\ \forall i<q$, (in all
$A$-bases).\\
Given the IB, one can determine the matrix $P(p)$ and the subset $S$ of
 $\real^q$ that represents the orbit space of the group action, as explained
 above.
 However, for an arbitrary compact coregular group,
 the IB is not known and difficult to determine. (For finite groups see
\cite{hum} and for simple Lie groups see \cite{sch}).\\
To classify the orbit spaces of all coregular compact linear groups it is
 possible to bypass the lack of knowledge of the IB's.
The many conditions found on the form of a general $P$-matrix and on the
 form of the complete factor $A$
suggest to try to find out all possible solutions to the boundary equation
 that are compatible with these {\it initial conditions}.
To be more precise, we only fix the dimension $q$ of $P(p)$ and we consider
 all matrix elements $P_{ab}(p)$ and $A(p)$ and $B(p)$ as unknown
 $w$-homogeneous polynomials satisfying the initial conditions.
In all these unknown polynomials the dependence on the variable $p_1$
can always be rendered explicit, even if all degrees $d_i$ are unknown.
We impose then the boundary equation and the condition $A(p)B(p) = \det(P(p))$
 to obtain a system of coupled differential equations that must be solved by
 $w$-homogeneous polynomial functions.
The initial conditions imposed are so strong that this system can be solved
 analytically and it gives only a finite number of different solutions for each
 value of the dimension $q$ (at least for $q$ = 2, 3, 4,
 where we have a complete list of solutions, but there is no
reason to believe that this will not be true for higher values of $q$).
At the end one obtains a list of all $P$-matrices satisfying the boundary
equation and the initial conditions.
 We called these matrices {\it allowable} $P$-matrices because they are
 potentially determined by an IB of an existing group $G$, but it is not
 known in general if that group does really exist. It is clear however that
 all $P$-matrices determined by the IB of the existing compact coregular
 groups are allowable $P$-matrices.
 The allowable $P$-matrices determine the subsets $S$ that can represent the
 orbit spaces of the compact coregular linear groups.\\
We worked out these very lenghty calculations and classified all allowable
 $P$-matrices
 of dimension 2, 3 and 4. They determine all possible orbit spaces of linear
 compact coregular groups of dimension 2, 3 and 4. Among our solutions there
 are all $P$-matrices corresponding to the irreducible finite group with up
to 4 basic invariants ($I_2(m),
 A_3, B_3, H_3, A_4, B_4, D_4, F_4, H_4$). They are easily singled out
 by looking at the degrees of the basic polynomials. To verify our results
the $P$-matrices of the groups listed above have also been determined
starting from the explicit form of the basic invariants \cite{4sv}.
 After a proper IBT all these $P$-matrices
fit exactly into our classification (but for a sign error found in the
coefficient of $p_1 p_3$ in the matrix element $P_{11}(p)$ of the solution $E5$
(corresponding to the group $H_4$) given in \cite{7st})\\

\section{The Solutions up to Dimension 4}

Here under I report a table of all 2-, 3- and 4-dimensional allowable
$P$-matrices, showing the corresponding degrees $[d_1,\ldots,d_{q-1},2]$
of the basic invariants and the degree $w(A)$ of the complete factor.
The parameters $j_i$ and $s$ that appear in the table are arbitrary
positive integers limited only by $d_1\geq \cdots \geq d_{q-1} \geq 2$.\\
The explicit forms of these allowable $P$-matrices are given in
 \cite{6st,7st}.\\
These allowable $P$-matrices share the following properties:
\begin{enumerate}
\item For each number $q$ of their dimension there are only a finite number of
 classes of solutions. Each class contains $P$-matrices that differ only in
 a scale factor $s$ that fixes the values of the degrees
 $[d_1,d_2,\ldots,d_q-1]$ and the powers of the variable $p_q$. In $\Pi$
all these matrices and the sections $S\cap\Pi$ become identical.
\item In convenient $A$-bases all coefficients of the $P$-matrices are
 integer numbers.
\item The subsets $S$ of $\real^q$ determined by the semipositivity of the
allowable $P$-matrices are connected.
\item All $P$-matrices that correspond to irreducible groups have
for complete factor $A(p)$ their whole determinant. In these cases
$A(p)$ always contains a term in $p_1^q$.
\end{enumerate}
We believe that these properties hold true for all values of $q$ and not only
 for $q \leq 4$.

\section{Induction Rules}

For dimensions $q\geq 5$ the analytic determination of the allowable
 $P$-matrices, in the same manner as it has been done for $q\leq 4$ asks
  for a discouraging amount of calculations.
Examining the allowable solutions that are available and chosing
 appropriately the $A$-bases, one finds many similarities between those of
 dimension 3 and those of dimension 4, especially:
\begin{enumerate}
\item on the explicit form of the matrix elements $P_{ab}(p)$;
\item on the explicit form of the complete factor $A(p)$;
\item on the geometrical shape of the sets $S$ that they determine in
 $\real^q$.
\end{enumerate}
These similarities lead to establish some induction rules that allow to
 write down easily most of the allowable $P$-matrices of dimension $(q+1)$
 out of those in dimension $q$.
Practically this can be done by adding, according to a given rule, a new
variable, and correspondingly a new row and column in the original $P$-matrix.

 Up to now 4 different types of induction rules have been found.
 Applying these rules to the solutions of dimension 2 and
 3 one can obtain all solutions of dimension 3 and 4 but the irreducible ones.
 These then appear as the only really new solutions.
It has been proved that the induction rules found are valid starting from any
 value of $q$, but it is likely that starting from $q\geq 4$ one may find new
 induction rules.
 The induction rules and their proofs are quite technical
 and will be presented elsewhere \cite{10st}.

\begin{table}[h]
\begin{center}
\caption{ALLOWABLE  $P$-MATRICES OF DIMENSION $q = 2, 3, 4$.}
\vskip 0.4cm
\begin{tabular}{|l|c|c|}
\hline
CLASS                     &      $w(A)$        &           $[d_1,\ldots,d_{q-1}]/s$     \\ \hline

$I$   (groups $I_2(s)$)   &     $ 2 d_1$       &           $    [1]$                    \\      \hline

$I(j_1,j_2)$              &   $  2 d_1 $       &    $    [(j_1+j_2)/2, 1]$              \\

$II(j_1)  $               &   $ 2 d_1+d_2 $    &     $     [j_1+1, 2]$                  \\

$III.1$   (group $A_3$)   &    $ 3 d_1   $     &   $        [4, 3]$                     \\

$III.2$   (group $B_3$)   & $   3 d_1   $      &   $        [6, 4]$                     \\

$III.3$   (group $H_3$)   & $     3 d_1  $     &   $       [10, 6]$                     \\     \hline

$A1(j_1,j_2,j_3,j_4)$     &    $  2 d_1   $    &$[(j_1+j_2)(j_3+j_4)/4, (j_1+j_2)/2, 1]$\\

$A3(j_1,j_2,j_3)$         &      $ 2 d_1+d_3  $& $   [(j_1+1)(j_2+j_3)/2, (j_1+1), 2] $ \\

$A2(j_1,j_2,j_3,j_4)$ &   $   2 d_1    $       & $  [(j_1+1)(j_2+j_3)/2+j_4, j_1+1, 2]$ \\

$A4(j_1,j_2)    $         & $ 2 d_1+j_1 d_3$   & $           [2j_2, j_1+j_2, 2]     $   \\

$A5(j_1,j_2,j_3)$         & $   2 d_1+d_2  $   & $ [j_1(j_2+1)+j_3, 2(j_2+1), 2]    $   \\

$A6(j_1,j_2,j_3)$         & $   2 d_1+d_2  $   & $    [(j_1+j_2)(j_3+1)/2, j_1+j_2, 1]$ \\

$A7(j_1,j_2)$             & $ 2 d_1+d_2+d_3$   & $        [j_1(j_2+1), 2j_1, 2]   $     \\

$A8(j_1,j_2)$             & $  2 d_1+2 d_2 $   &$          [j_1+1, j_2+1, 2]     $      \\

$B1(j_1)    $             & $   2 d_1     $    & $         [6j_1, 4, 3]           $     \\

$B2         $             & $  3 d_1      $    & $         [4, 3, 3]              $     \\

$B3(j_1,j_2)$             & $    3 d_1    $    & $   [2(j_1+j_2), 3(j_1+j_2)/2, 1]$     \\

$B4(j_1)    $             & $ 3 d_1+d_3   $    & $        [4j_1, 3j_1, 2]        $      \\

$C1(j_1,j_2)$             & $    2 d_1    $    & $      [3(j_1+2j_2), 6, 4]      $      \\

$C2(j_1)    $             & $ 2 d_1+d_2   $    & $         [6j_1, 6, 4]          $      \\

$C3(j_1)    $             & $ 2 d_1+2 d_2 $    & $       [3(j_1+1), 6, 4]        $      \\

$C4         $             & $  3 d_1      $    & $        [6, 4, 3]              $      \\

$C5(j_1,j_2)$             & $    3 d_1    $    & $   [3(j_1+j_2), 2(j_1+j_2), 1] $      \\

$C6(j_1)    $             & $ 3 d_1+d_3   $    & $        [6j_1, 4j_1, 2]        $      \\

$D1(j_1)    $             & $    2 d_1    $    & $        [15j_1, 10, 6]         $      \\

$D2         $             & $   3 d_1     $    & $         [10, 6, 4]            $      \\

$D3(j_1,j_2)$             & $    3 d_1    $    & $   [5(j_1+j_2), 3(j_1+j_2), 1] $      \\

$D4(j_1) $                & $  3 d_1+d_3  $    & $         [10j_1, 6j_1, 2]   $         \\

$E1$      (group $A_4$)   &   $  4 d_1  $      &$            [5, 4, 3]       $          \\

$E2$      (group $D_4$)   &   $ 4 d_1   $      &$            [6, 4, 4]      $           \\

$E3$      (group $B_4$)   &   $ 4 d_1   $      &$            [8, 6, 4]      $           \\

$E4$      (group $ F_4$)  &  $  4 d_1  $       &$           [12, 8, 6]      $           \\

$E5$      (group $ H_4$)  &  $  4 d_1  $       &$          [30, 20, 12]     $            \\ \hline
\end{tabular}
\end{center}
\end{table}

\section{Conclusions}

The main conclusions of our calculations are as follows:
\begin{enumerate}
\item It is possible (besides computational difficulties) to classify all
 allowable $P$-matrices. They determine univocally the sets $S$ that can
 be identified with the orbit spaces of compact coregular linear groups.
\item The determination of the allowable $P$-matrices is done without assuming
 a specific integrity basis and without knowing any specific information of
 group structure, but using only some very general algebraic conditions.
\item It is not proved if it does always exist a compact coregular group
 corresponding to a given allowable $P$-matrix but the converse is true, that
 is, all existing compact coregular groups determine $P$-matrices of the same
 form (perhaps after an IBT) of an allowable $P$-matrix. The classification
 of the allowable $P$-matrices then
 gives at least a selection rule for the degrees of the basic invariants
 of the compact coregular linear groups.
\item The explicit form of the $P$-matrices is strongly related to the form of
 the basic invariants and may help to determine the IB in all those cases when
 it is unknown.
\end{enumerate}
The main open problems in all this subject are the following:
\begin{enumerate}
\item Does a compact coregular linear group corresponding to
 an arbitrary allowable $P$-matrix always exist? If this group exists, which
 is the group and what is its integrity basis?
\item What is the meaning of the induction rules and what is their relation
 with group theory?
\item Why can all allowable $P$-matrices be expressed, in some proper IB,
 with only integer numbers for coefficients of the variables $p_i$?
\item Do the $P$-matrices corresponding to irreducible groups always have
 their active factor $A(p)$
 coincident with $det(P(p))$ and containing a term in $p_1^q$?
\end{enumerate}
Our results are partial but they point out a very strong relation with
 classical group theory and with invariant theory which ought to be further
 investigated.

\end{document}